\documentclass[12pt,preprint]{aastex}
\usepackage{emulateapj5,psfig,apjfonts}

\let\lsim=\la
\let\gsim=\ga
\newcommand\deltasigma{\delta_\Sigma}

\newcommand\deltal{\delta_{\rm L}}
\newcommand\Vsurvey{V_0}
\newcommand\Vsphere{V_{\rm S}}

\newcommand\NB{{\rm NB711}}

\slugcomment{ApJL, accepted}

\shorttitle{Large-Scale Structure at Redshift $\simeq 5$}
\shortauthors{Shimasaku et al.}

\begin{document}


\title{Subaru Deep Survey. IV.
Discovery of a Large-Scale Structure 
at Redshift $\simeq 5$\altaffilmark{1} }


\author{K. Shimasaku   \altaffilmark{2,3},
M. Ouchi      \altaffilmark{2},
S. Okamura    \altaffilmark{2,3},
N. Kashikawa  \altaffilmark{4},
M. Doi        \altaffilmark{5},\\
H. Furusawa   \altaffilmark{6},
M. Hamabe     \altaffilmark{7},
T. Hayashino  \altaffilmark{8},
K. Kawabata   \altaffilmark{4},
M. Kimura     \altaffilmark{9},\\
K. Kodaira    \altaffilmark{10},
Y. Komiyama   \altaffilmark{6},
Y. Matsuda    \altaffilmark{8},
M. Miyazaki   \altaffilmark{2},
S. Miyazaki   \altaffilmark{6},\\
F. Nakata     \altaffilmark{11},
K. Ohta       \altaffilmark{12},
Y. Ohyama     \altaffilmark{6},
M. Sekiguchi  \altaffilmark{9},
Y. Shioya     \altaffilmark{13},\\
H. Tamura     \altaffilmark{8},
Y. Taniguchi  \altaffilmark{13},
M. Yagi       \altaffilmark{4}, 
T. Yamada     \altaffilmark{4},
and N. Yasuda     \altaffilmark{4}
}

\email{shimasaku@astron.s.u-tokyo.ac.jp}


\altaffiltext{1}{Based on data collected at Subaru Telescope, 
which is operated by the National Astronomical Observatory of Japan.}
\altaffiltext{2}{Department of Astronomy, School of Science,
        University of Tokyo, Tokyo 113-0033, Japan; 
        shimasaku@astron.s.u-tokyo.ac.jp}
\altaffiltext{3}{Research Center for the Early Universe, 
        School of Science,
        University of Tokyo, Tokyo 113-0033, Japan}
\altaffiltext{4}{National Astronomical Observatory of Japan, 
Mitaka, Tokyo 181-8588, Japan}
\altaffiltext{5}{Institute of Astronomy, School of Science, 
University of Tokyo, Mitaka, Tokyo 181-0015, Japan}
\altaffiltext{6}{Subaru Telescope, 
National Astronomical Observatory of Japan, 
650 N. A'ohoku Place, Hilo, HI 96720, USA}
\altaffiltext{7}{Department of Mathematical and Physical Sciences,
Faculty of Science, Japan Women's University, Tokyo 112-8681, Japan}
\altaffiltext{8}{Research Center for Neutrino Science,
Graduate School of Science, Tohoku University,
Aramaki, Aoba, Sendai 980-8578, Japan}
\altaffiltext{9}{Institute for Cosmic Ray Research, 
University of Tokyo, Kashiwa, Chiba 277-8582, Japan}
\altaffiltext{10}{The Graduate University for Advanced Studies 
(SOKENDAI), Shonan Village, Hayama, Kanagawa 240-0193, Japan}
\altaffiltext{11}{Department of Physics, University of Durham,
South Road, Durham DH1 3LE, UK}
\altaffiltext{12}{Department of Astronomy, Kyoto University,
Sakyo-ku, Kyoto 606-8502, Japan}
\altaffiltext{13}{Astronomical Institute, Graduate School of Science,
Tohoku University, Aramaki, Aoba, Sendai 980-8578, Japan}


\begin{abstract}
We report the discovery of a large-scale structure 
of Lyman $\alpha$ emitters (LAEs) at $z=4.86$ 
based on wide-field imaging with the prime-focus camera 
(Suprime-Cam) on the Subaru telescope.
We observed a $25' \times 45'$ area of the Subaru Deep Field 
in a narrow band 
(\NB, $\lambda_{\rm C}=7126$\AA\hspace{3pt}and FWHM$=73$\AA) 
together with $R$ and $i'$.
We isolate from these data 43 LAE candidates down to $\NB = 25.5$ 
mag using color criteria.
Follow-up spectroscopy of five candidates suggests 
the contamination by low-$z$ objects to be $\sim 20\%$.
We find that the LAE candidates are clustered in an elongated region 
on the sky of 20 Mpc in width and 50 Mpc in length at $z=4.86$, 
which is comparable in size to present-day large-scale structures
(we adopt $H_0=70$ km s$^{-1}$ Mpc$^{-1}$, 
$\Omega_0=0.3$, and $\lambda_0=0.7$).
This elongated region includes
a circular region of 12 Mpc radius of higher surface overdensity 
($\deltasigma=2$), 
which may be the progenitor of a cluster of galaxies.
Assuming this circular region to be a sphere with a spatial 
overdensity of 2, we compare our observation with 
predictions by Cold Dark Matter models.
We find that an $\Omega_0=0.3$ flat model with $\sigma_8=0.9$ 
predicts the number of such spheres consistent with 
the observed number (one sphere in our survey volume) 
if the bias parameter of LAEs is $b \simeq 6$.
This value suggests that the typical mass of dark haloes hosting 
LAEs at $z\simeq 5$ is of the order of $10^{12} M_\odot$. 
Such a large mass poses an interesting question about the nature
of LAEs.

\end{abstract}


\keywords{cosmology: observations ---
          cosmology: early universe ---
          cosmology: large-scale structure of universe ---
          galaxies: high-redshift ---
          galaxies: evolution ---
          galaxies: photometry }


%
%

\section{INTRODUCTION}

The formation and evolution of the large-scale clustering of galaxies, 
as seen in the present-day universe, are central issues in cosmology.
Observations of galaxy clustering in the distant universe 
give us great clues to this issue, including how galaxies 
were formed in the underlying dark matter.
To date, many efforts have been made to detect large-scale 
clustering of galaxies at high redshifts.
Probable detections of galaxy clusters on 
Megaparsec scales have been reported 
(e.g., Giavalisco et al. 1994;
Pascarelle et al. 1996a; 
Le F\`evre et al. 1996;
Malkan, Teplitz, \& McLean 1996;
Francis et al. 1996;
Keel et al. 1999;
Campos et al. 1999;
Pentericci et al. 2000).
The most remarkable observation is 
the discovery of a high concentration of Lyman-break Galaxies 
(LBGs) at $z\sim 3$ with a size of at least $11' \times 8'$
(or $21 \times 15$ Mpc) by Steidel et al. (1998), 
who argue that LBGs must be very biased tracers of mass 
if such a structure is consistent with Cold Dark Matter (CDM) models.
More recently, Venemans et al. (2002) have discovered 
a proto-cluster of $M \sim 10^{15} M_\odot$ 
with a size of $2.7 \times 1.8$ Mpc around a radio galaxy at $z=4.1$ 
through a narrow-band survey.

To place stringent constraints on models of structure formation, 
it is crucial to observe galaxy clustering over a wide 
range of scales at various epochs.
In this {\it Letter}, we report on a narrow-band survey 
in a $25' \times 45'$ field for Lyman $\alpha$ Emitters (LAEs) 
at $z=4.86$.
Our main goal is 
to search at high redshifts for the progenitors of 
present-day large-scale structures like the Great Wall 
(Geller \& Huchra 1989), in order to trace the evolution of 
structures on scales of tens of Mpc. 
We also aim at detecting (proto) clusters at $z \sim 5$.
Unless otherwise noted, 
we adopt $\Omega_0=0.3$ and $\lambda_0=0.7$, 
and express the Hubble constant as 
$H_0 = 70 h_{70}$ km s$^{-1}$ Mpc$^{-1}$.

%
%

\section{OBSERVATIONS}

\subsection{Imaging}

We carried out a deep imaging survey in the sky area of 
the Subaru Deep Field 
(SDF; centered at $(13^h 24^m 21.^s4, +27^\circ 29' 23'')$ 
[J2000.0]; Maihara et al. 2001) 
in the $R$ and $i$ bands and a narrow-band 
filter centered at 7126 \AA, \NB,
with the prime-focus camera (Suprime-Cam; Miyazaki et al. 2002) 
on Subaru in March - June 2001 and May 2002.
The FWHM of the \NB\hspace{2pt} filter is 73 \AA 
\footnote{The central wavelength and the FWHM of the \NB\hspace{2pt} 
filter are the values measured for the F/1.86 prime focus.}, 
giving a survey depth for LAEs along the sightline of 
$\Delta z = 0.06$, or equivalently $33 h_{70}^{-1}$ Mpc.
We observed two field-of-views (FoVs) of Suprime-Cam 
allowing for a wide overlap: 
the central FoV and the northern FoV
\footnote{We first observed the central FoV and found 
in a northern part
a large-scale overdense region of LAE candidates.
We then imaged the northern FoV with an overlap of $15'$ 
to check the reliability of this overdense region
as well as to trace a northern extension of the overdense region.
We found consistently an overdense region 
in the data of the northern FoV. 
The match of LAE candidates in the overlapped region 
between the two FoV data is $\simeq 60\%$, which can be regarded as 
an estimate of the completeness of our LAE detection.}.

\centerline{\psfig{file=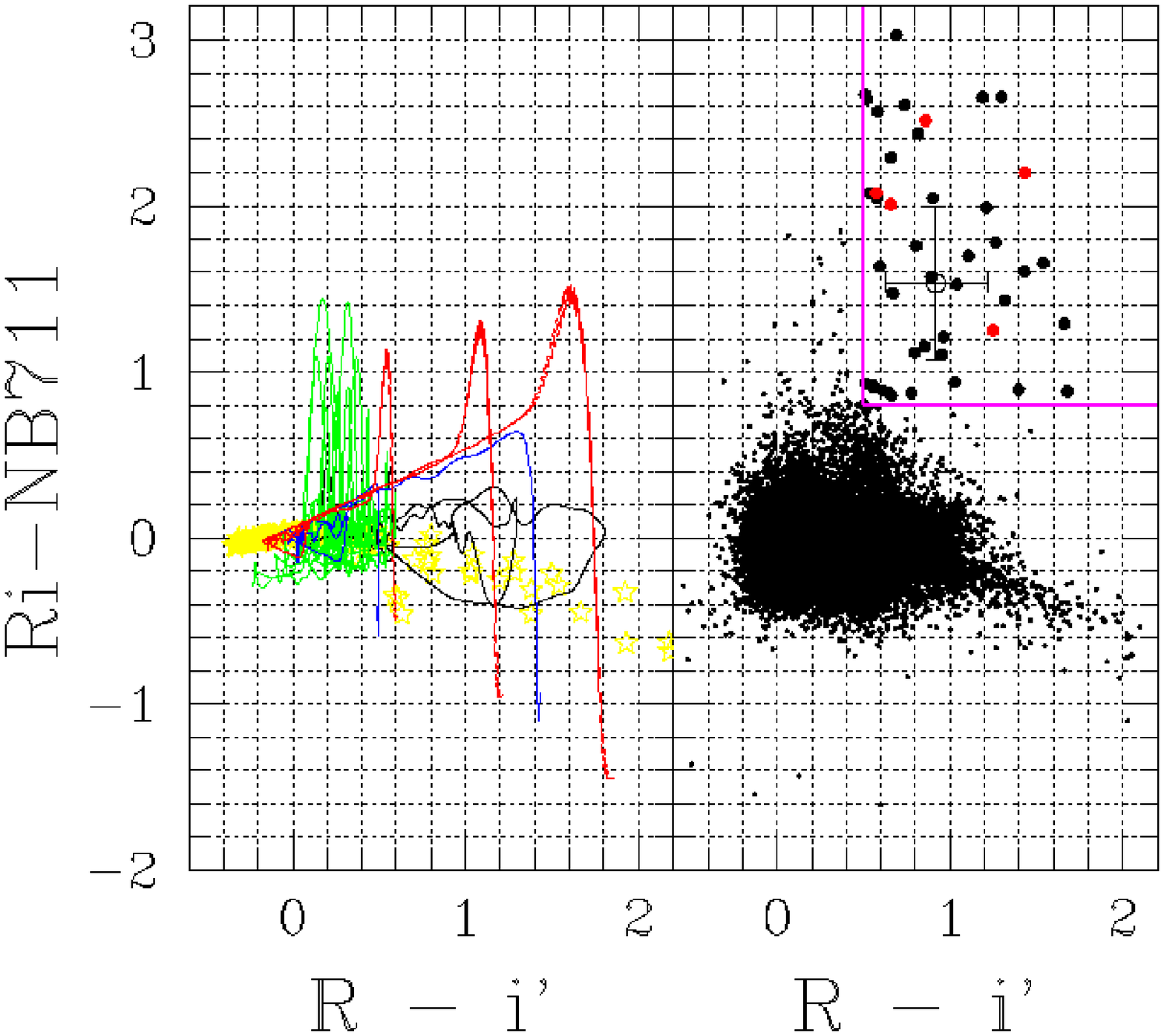,width=3.0in}}
\noindent{\scriptsize \addtolength{\baselineskip}{-3pt}
{\sc Fig.~1.} ---
Two-color diagrams for continuum color ($R-i'$)
and narrow-band excess color ($Ri-\NB$), where
$Ri \equiv (R+i)/2$ is a continuum magnitude.
({\it left panel})
Tracks of model galaxies at different redshifts.
Red lines indicate model LAEs which are
a composite spectrum of a 0.03 Gyr single burst
model galaxy (GISSEL96; Bruzual \& Charlot 1993)
and a Lyman $\alpha$ emission
($EW_{\rm rest}$(Ly$\alpha$)=22 \AA);
from the left to right, three different amplitudes
of IGM absorption are applied:
$0.5\tau_{\rm eff}$, $\tau_{\rm eff}$, and
$1.5\tau_{\rm eff}$, where $\tau_{\rm eff}$ is
Madau's (1995) median opacity.
The narrow-band excess in each of the peaks in
the red lines indicates the Lyman $\alpha$ emission
of LAEs at $z=4.86$.
Green lines show six templates of
nearby starburst galaxies, with different dust extinction 
($E(B-V)=0.0-0.7$), given in Kinney et al. (1996) 
redshifted up to $z=1.2$.
The narrow-band excess peaks in the green lines 
correspond to the emission line of
H$\alpha$ ($z=0.08$), [OIII]($z=0.4$), H$\beta$($z=0.5$),
or [OII]($z=0.9$).
Two blue lines indicate the tracks for model
LBGs without Lyman $\alpha$ emission, with 
$0.5\tau_{\rm eff}$ and $\tau_{\rm eff}$, respectively.
Black lines show colors of typical
elliptical, spiral, and irregular galaxies 
of Coleman, Wu, \& Weedman (1980) 
which are redshifted from $z=0$ to $z=2$. 
Yellow star marks show 175 Galactic stars given by 
Gunn \& Stryker (1983), and 24 Kurucz model stars.
({\it right panel}) Colors of the detected objects.
Large circles indicate 43 LAE candidates, 
among which red ones have spectroscopic observations, 
while small dots are for the other objects.
An open circle with error bars indicates the median colors
and their errors for the 43 LAE candidates.
Our LAE selection criteria are outlined by a pink line.
}
\medskip

The data of the central FoV have already been reduced and used 
to study general properties of LAEs by Ouchi et al. (2003a).
For the northern FoV, individual CCD data were reduced and
combined using IRAF and our own data reduction software 
(Yagi et al. 2002).
For the overlapped region, we use the data of the central FoV.
The total exposure time and the limiting magnitude 
($3\sigma$ on a $1.''8$ aperture) are: 
90 min and 27.1 mag ($R$), 
138 min and 26.9 mag ($i'$), and
162 min and 26.0 mag (\NB) for the central FoV; 
120 min and 27.5 mag ($R$), 
110 min and 27.3 mag ($i'$), and
144 min and 26.1 mag (\NB) for the northern FoV. 
All magnitudes are AB magnitudes.
The seeing sizes of the final images are $0.''90$.

Object detection and photometry are 
made using SExtractor version 2.1.6 (Bertin \& Arnouts 1996).
The \NB-band image is chosen to detect objects. 
In this {\it Letter}, 
to ensure secure photometry and a low contamination 
for our LAE sample, we confine ourselves to objects 
brighter than \NB=25.5 (their number is $34,653$).
We apply the following criteria (Ouchi et al. 2003a) to these objects,
to isolate LAEs at $z = 4.86$:
$Ri - \NB > 0.8$, $R - i > 0.5$, and $i'-\NB>0$, 
where $Ri \equiv (R+i')/2$.
The second criterion reduces contamination by foreground galaxies 
whose emission lines other than Lyman $\alpha$ 
happen to enter the $\NB$ band (see Fig. 1). 
The number of objects passing the above criteria is 43. 
Figure 1 plots on the $Ri-\NB$ vs $R-i'$ plane 
all objects with $\NB\le25.5$ (right) 
and model galaxies and Galactic stars (left).
This figure demonstrates that the second criterion efficiently 
removes low-$z$ objects but selects LAEs at $z = 4.86$
(Ouchi et al. 2003a).

\vspace{-70pt}
\centerline{\psfig{file=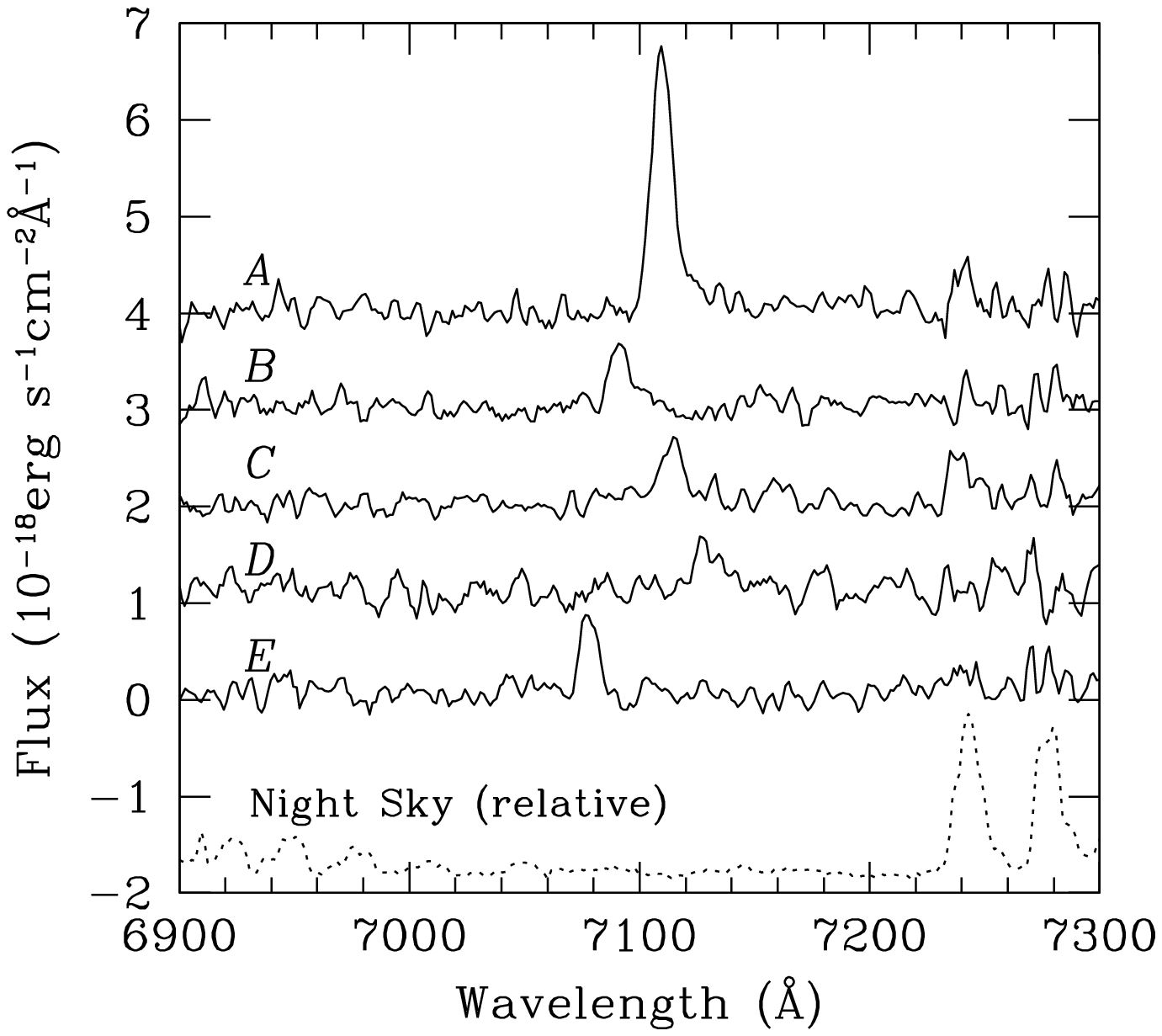,width=4.0in}}
\noindent{\scriptsize \addtolength{\baselineskip}{-3pt}
{\sc Fig.~2.} ---
Spectra of five candidates over the range 6900-7300\AA.
Flux of objects $A$, $B$, $C$, and $D$, has been offset 
for clarity by 
$4\times10^{-18}$, 
$3\times10^{-18}$, 
$2\times10^{-18}$, 
and $1\times10^{-18}$ 
ergs s$^{-1}$ cm$^{-2}$ \AA$^{-1}$, respectively.
The (relative) spectrum of the night sky emission is shown by 
a dotted line; emission lines are not present around 7100\AA.
}
\medskip

%
%

\subsection{Spectroscopy}

We selected a $6'\phi$ field of the highest overdensity 
(see \S 3.1) for multi-slit spectroscopy, 
and made spectroscopic observations for five candidates 
in this field with FOCAS (Kashikawa et al. 2002) 
on Subaru on June 6, 2002, to check our photometric selection.
Table 1 summarizes their properties.
The exposure time was 2 hr for each object. 
The seeing size was $0.''7-0.''8$.
We used the 300 line/mm grating with a dispersion of 
1.4 \AA\hspace{3pt}pixel$^{-1}$
and a wavelength coverage of 4700-9400 \AA.
We adopted a slit width of $0.''8$, which gave  
a spectral resolution of 9.8 \AA.
The continuum flux limit of our spectroscopy was 
$6.3 \times 10^{-19}$ erg s$^{-1}$ cm$^{-2}$ \AA$^{-1}$ 
($5\sigma$).

Figure 2 shows spectra of the five candidates 
after two-pixel smoothing.
All the spectra have an emission line near 7100\AA. 
Possible candidates (other than Lyman $\alpha$) 
for this emission line are 
H$\alpha$, H$\beta$, [OIII]$\lambda 5007$, and [OII]$\lambda 3727$.
We find that the lines are not [OIII]$\lambda 5007$ at $z\simeq 0.4$ 
because of the lack of the H$\beta$ line at the corresponding 
wavelength;
from similar reasons, they are not either H$\alpha$ or H$\beta$.
Thus, the five objects are either an LAE or an [OII] emitter.

\medskip

\centerline{
Table 1. Five candidates with spectroscopic observations.
}

\centerline{
\begin{tabular}{cccccrl}
\hline
ID  &  $R^{(a)}$  & $i'^{(a)}$ & \NB$^{(a)}$ & $z$ & 
    EW$^{(b)}$ &
    $f^{(c)}$ \\
 & (mag) & (mag) & (mag) &      &   (\AA)   &      \\
\hline
$A$ & 27.18 & 26.32 & 24.24 & 4.850 &  51 & 3.5  \\
$B$ & 28.25 & 26.82 & 25.33 & 4.834 &  15 & 0.77 \\
$C$ & 27.22 & 26.56 & 24.88 & 4.856 &  14 & 0.93 \\
$D$ & 27.21 & 25.96 & 25.33 & 4.865 &   6 & 0.49 \\
$E$ & 27.62 & 27.05 & 25.26 & 4.825$^{(d)}$ &  18 & 1.1  \\
\hline
\end{tabular}
}

\noindent{\scriptsize \addtolength{\baselineskip}{0pt}
(a) AB magnitudes.
}

\noindent{\scriptsize \addtolength{\baselineskip}{0pt}
(b) Rest-frame equivalent width (\AA) 
estimated from the spectrum, assuming the line to be Lyman $\alpha$.
Due to very weak continuum emissions and a relatively 
short exposure of spectroscopy, EW values should have 
large errors.
}

\noindent{\scriptsize \addtolength{\baselineskip}{0pt}
(c) Line flux in units of 
$10^{-17}$ erg s$^{-1}$ cm$^{-2}$.
We have not examined the profiles of Lyman $\alpha$ emissions
because of low S/N ratios of the spectra, 
although $A$ and possibly $B$ appear to
have wide velocity widths, which may suggest starburst winds.
}

\noindent{\scriptsize \addtolength{\baselineskip}{0pt}
(d) $z=0.90$ if the line is [OII].
}

\medskip

\vspace{-5pt}
\centerline{\psfig{file=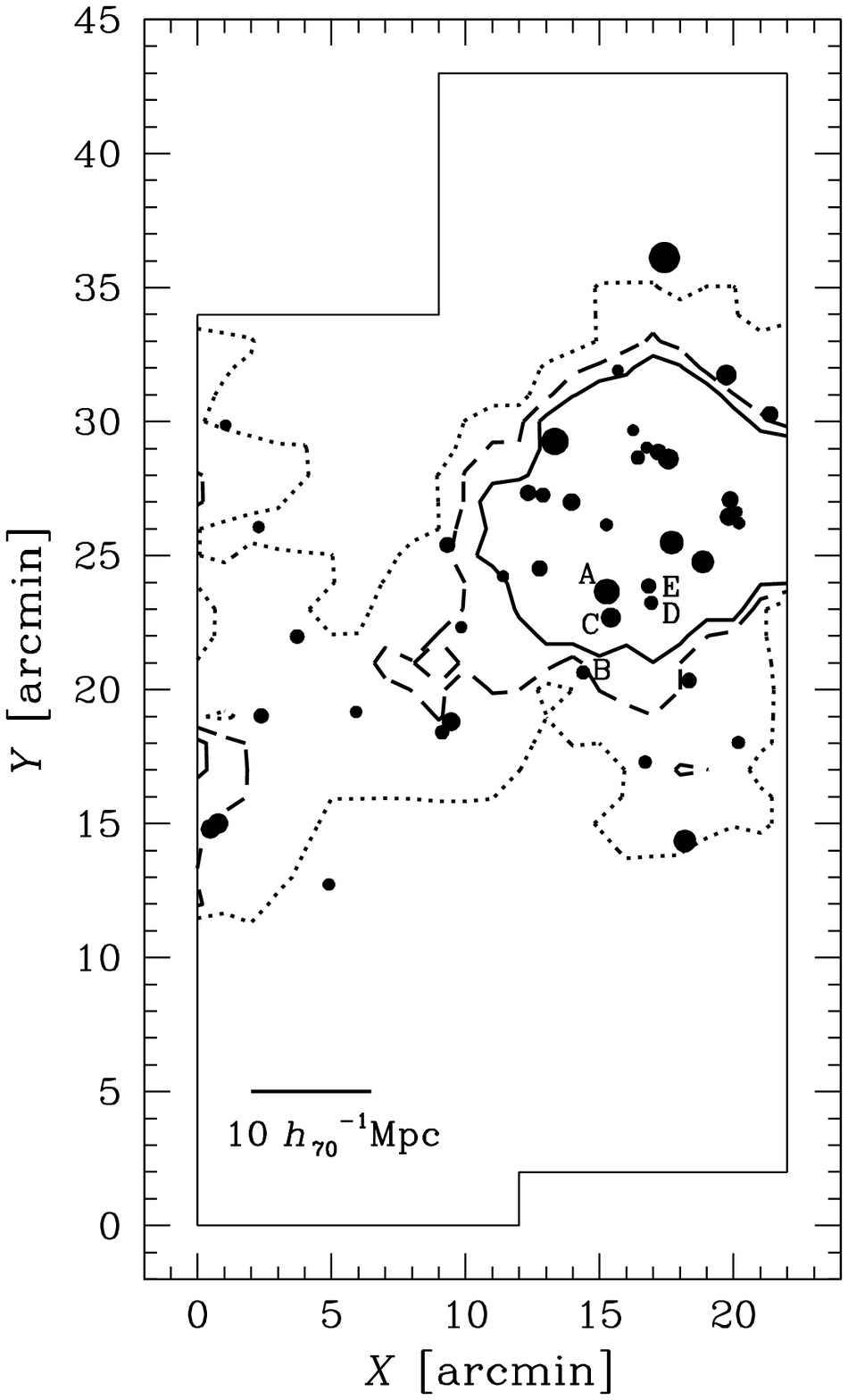,width=4.0in}}
\noindent{\scriptsize \addtolength{\baselineskip}{-3pt}
{\sc Fig.~3.} ---
Sky distribution of 43 LAE candidates.
North is up and east is to the left.
Areas of relatively poor quality have been trimmed.
A bright star in the upper left corner, around which 
detection of LAEs is impossible, has been masked.
LAE candidates are shown by circles. 
Brighter candidates are shown by larger symbols.
Dotted, dashed, and solid lines correspond to contours of 
$\deltasigma=0, 1$, and 2, respectively
(A top-hat smoothing of $8 h_{70}^{-1}$ Mpc radius 
is made over the LAE distribution to compute the local overdensity).
The region with $\deltasigma \ge 2$ can be approximated 
by a circle of $12 h_{70}^{-1}$ Mpc radius.
}
\medskip

We conclude from the following reasons that three 
($A$, $B$, $D$) are convincing LAEs and $C$ is a probable LAE 
while $E$ may be an [OII] emitter at $z\sim 0.9$.
First, the emission line of $A$ and $B$ shows an asymmetric shape 
with a blueward cutoff; this is a common feature of 
high-redshift Lyman $\alpha$ emission (e.g., Dey et al. 1998).
Second, although we cannot rule out the possibility 
of $D$ being an [OII] emitter on the basis of its spectrum alone, 
we have seen in Figure 2 that its $R-i'$ color is 
red enough to match the expectation for LAEs.
Moreover, Ouchi et al. (2003b) find that $C$ and $D$ satisfy 
the selection criteria for LBGs at $z=4.7 \pm 0.5$ 
in the $V-i'$ vs $i'-z'$ plane
\footnote{The five candidates are located 
in the central part of the SDF, 
where we obtained deep $B,V,z'$ images as well 
in March - June 2001 (Ouchi et al. 2003b).}.
Indeed, the $V-i'$ colors of $A,B,C,D$ are $\sim 1.7-2.5$, 
being consistent with the color expected for $z\sim 5$ galaxies; 
on the other hand, the color of $E$ is $\sim 0.5$.
From the above discussion, 
we estimate the contamination of our sample to be 
$\sim 1/5=20\%$ 
\footnote{
This value is lower than 
that estimated by Ouchi et al. (2003a) for a deeper sample, 
$\sim 40\%$, on the basis of photometric properties alone. 
This is probably because the photometric errors in our sample 
are smaller than those in Ouchi et al.}.

As an independent check, 
we have examined the distribution of objects 
with $Ri - \NB > 0.8$ but $R - i < 0.4$, 
which are probably low-$z$ interlopers, and have found that 
their distribution on the sky is almost uniform 
and does not correlate with that of the LAE candidates.

\vspace{-95pt}
\centerline{\psfig{file=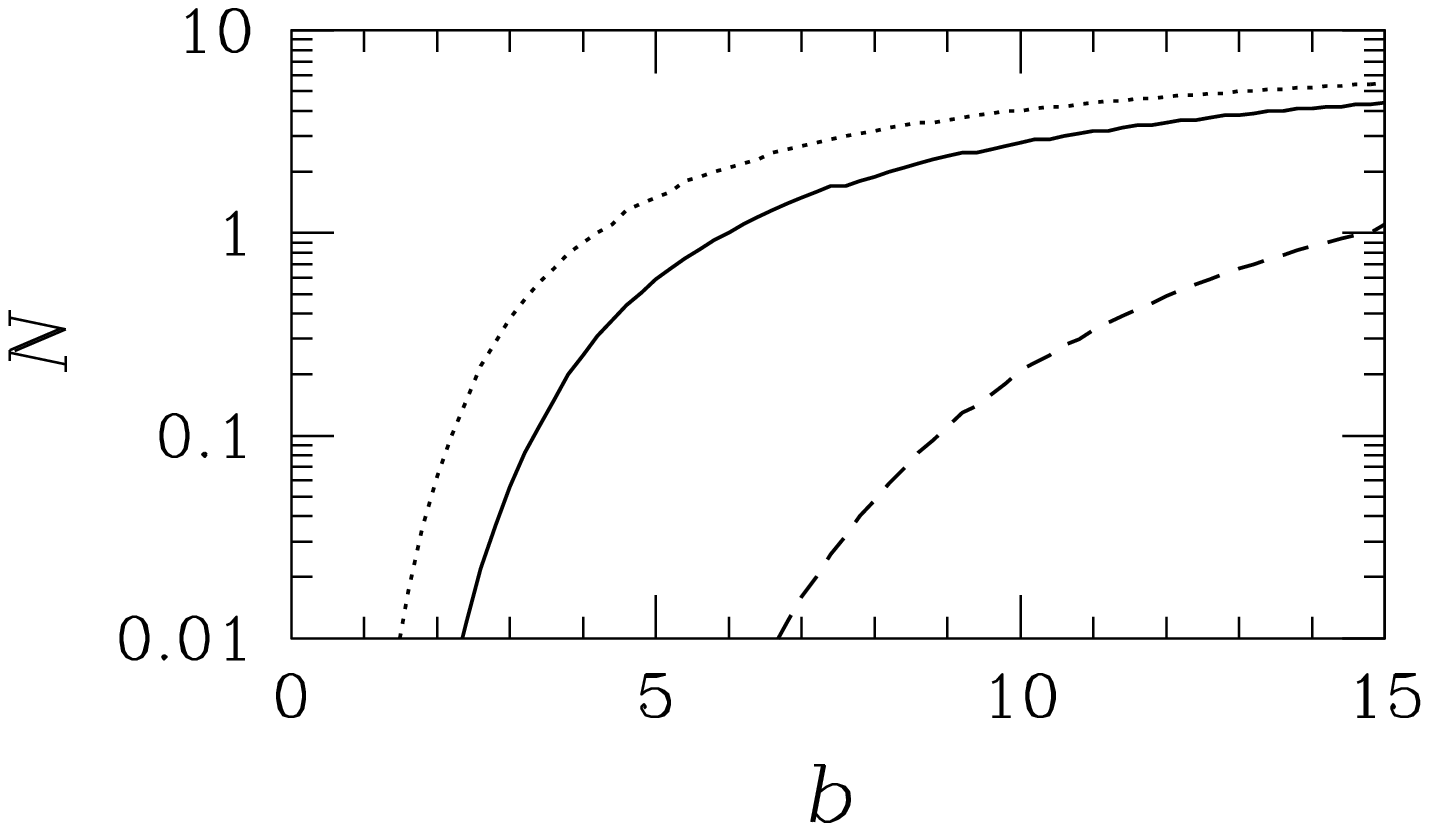,width=4.0in}}
\vspace{-50pt}
\noindent{\scriptsize \addtolength{\baselineskip}{-3pt}
{\sc Fig.~4.} ---
Number of $\delta\ge2$ spheres in our survey volume 
predicted by three CDM models.
Solid, dotted, and dashed lines indicate 
the $\Lambda$-dominated, open, and standard CDM models, respectively.
}
\medskip

%
%

\section{RESULTS AND DISCUSSION}

\subsection{Large-Scale Structure of LAEs}

Figure 3 shows the sky distribution of 
photometrically selected 43 LAE candidates.
We find a remarkable, large-scale clustering; 
an overdense region lies in the ESE - WNW direction 
and there are few objects outside this region.
To quantify overdensity, 
we estimate the local surface density of LAEs, 
$\Sigma(x,y)$, and compute the surface overdensity 
$\deltasigma(x,y) \equiv [\Sigma(x,y)-\bar{\Sigma}]/\bar{\Sigma}$, 
where $\bar{\Sigma}$ is the mean surface density of LAEs.
We adopt as $\bar{\Sigma}$ the mean surface density in our image, 
although a larger area is desirable to obtain a more accurate 
(i.e., global) estimate of $\bar{\Sigma}$.

The overdensity contours are drawn in Fig. 3.
The projected size of the overdense region 
($\deltasigma \ge 0$) is found to be about 20 $h_{70}^{-1}$Mpc 
in width and larger than 50 $h_{70}^{-1}$Mpc in length 
(comoving units)
since the region seems to 
continue outside the image at either side.
This elongated overdense structure may be 
a cross section of a wall-like  
structure which extends along the sightline 
(Note that the survey depth is $33 h_{70}^{-1}$ Mpc). 
Another possibility may be that we are seeing a {\lq}redder{\rq} part 
of a structure which is centered at a redshift smaller 
than that corresponding to the center of \NB ($z=4.86$), 
since all five objects with spectroscopy have $z \lsim 4.86$.
This elongated structure includes a region of 
$\deltasigma \ge 2$ which is approximated well 
by a circle of 12 $h_{70}^{-1}$Mpc radius.
Since the minimum mass of this circular region is computed to be 
$\simeq (4\pi/3)\rho_0 12^3 \simeq 3\times 10^{14} h_{70}^{-1} M_\odot$ 
($\rho_0$ is the mean matter density of the universe), 
it may become later a massive cluster of galaxies 
after collapsing to a size of a few Mpc.
In the present-day universe, clusters are often embedded in 
large-scale structures. 
From these features, it is very likely that this elongated region 
is a proto large-scale structure at $z\simeq 5$.

%
%

\subsection{Implications on Cold Dark Matter Models}

We examine in a simple manner whether or not CDM models 
can reproduce the observed structure.
We focus on the circular region of $\deltasigma \ge 2$, 
since properties of circular regions are predicted easily.
We make a reasonable assumption that this region 
is a sphere of 12 $h_{70}^{-1}$Mpc radius in 3-dimensional space 
with a spatial overdensity of $\delta=2$. 
We consider three CDM models: 
a $\Lambda$-dominated model with $\Omega_0=0.3$, $\lambda_0=0.7$,
$H_0 = 70$ km s$^{-1}$ Mpc$^{-1}$, $\sigma_8=0.9$; 
an open model with $\Omega_0=0.3$, $\lambda_0=0$,
$H_0 = 70$ km s$^{-1}$ Mpc$^{-1}$, $\sigma_8=0.9$; 
the standard model with $\Omega_0=1$, $\lambda_0=0$, 
$H_0 = 50$ km s$^{-1}$ Mpc$^{-1}$, $\sigma_8=0.5$. 
We adopt these $\sigma_8$ values from 
the $\sigma_8$ - $\Omega_0$ relation derived by 
Eke, Cole, \& Frenk (1996) for local X-ray clusters.

Using linear perturbation theory, 
we calculate the number of spheres within our observed 
volume above $\delta = 2$ as (cf. Steidel et al. 1998):
\begin{equation}
N = {\Vsurvey \over{\Vsphere}} 
\int_{\deltal}^\infty 
  {1\over{\sqrt{2\pi}\sigma_z}}
  \exp\left[ - {x^2\over{2\sigma_z^2}} \right] dx,
\end{equation}
\noindent
where $\Vsurvey$ is the volume surveyed by our observation 
($1.4\times 10^5 h_{70}^{-3}$ Mpc$^3$), 
$\Vsphere$ is the volume of an unperturbed sphere 
whose mass is equal to 
$(1+2/b){4\pi\over{3}}\rho_0 (12 h_{70}^{-1})^{3}$, 
$\deltal$ is the linearly extrapolated mass overdensity 
estimated from 
the observed mass overdensity 
(we use a fitting-formula in Bernardeau 1994), 
$\sigma_z$ is the rms fluctuation of mass overdensity 
at $z=4.86$ on top-hat filter on a scale of 
the unperturbed sphere, 
and $b$ is the linear bias parameter for LAEs. 

Figure 4 plots $N$ against $b$ for the three models. 
We find that the best-fit value of $b$ that produces 
the observed number ($=1$) of spheres is $b \simeq 6$ 
($\Lambda$-dominated model), 
$\simeq 4$ (open), and
$\simeq 15$ ($\Omega_0=1$).
We can also estimate the lower and upper 
limits of $b$ by imposing $0.05 \le N \le 4.7$ 
($95\%$ C.L. assuming Poisson statistics); 
we obtain $b\sim 3-16$, $b\sim 2-12$, and $b\sim 8-37$ 
for the $\Lambda$-dominated, open, and $\Omega_0=1$ models. 
Thus, LAEs at $z\simeq 5$ must be highly biased tracers of mass.
If we use the linear bias model of Mo \& White (1996)  
which relates the mass of a dark halo with its $b$, 
we obtain, from the best-fit $b$ values, 
the mass of dark haloes hosting our LAEs to be 
$\sim 1 \times 10^{12}$, 
$2 \times 10^{12}$, and 
$2 \times 10^{12} h_{70}^{-1} M_\odot$  
for the $\Lambda$-dominated, open, and $\Omega_0=1$ models, 
respectively.

The best-fit values of $b$ we find for our LAE candidates 
are similar to or possibly higher than those for LBGs at $z\sim 3-4$ 
obtained in the following work.
From an analysis of a galaxy overdensity 
of $\sim 10^{15} M_\odot$ at $z=3.09$, 
Steidel et al. (1998) have estimated 
$b \gsim 4$ for the $\Omega_0=0.3$ flat model.
Analyses of the two-point correlation function have given 
$b \sim 2-3$ for LBGs at $z\sim 3$ and $4$ 
for an $\Omega_0=0.3$ flat model  
with the same normalization as we adopt
(Porciani \& Giavalisco 2002; Ouchi et al. 2001).
Higher $b$ at $z\approx 5$ would be reasonable if we recall that 
$b$ for a given halo mass is predicted to increase with $z$.
Indeed, Ouchi et al. (2003b) obtain $b \sim 6$ for 
LBGs at $z\sim 5$, comparable to our value.

Steidel et al. (1998) and Porciani \& Giavalisco (2002) have 
inferred the mass of haloes hosting LBGs 
to be of the order of $10^{12} M_\odot$. 
The $b$ value for $z\sim 5$ LBGs estimated by Ouchi et al. (2003b) 
also gives a similar mass.
Hence, haloes hosting $z\simeq 5$ LAEs 
are roughly as massive as those of LBGs at $z\sim 3-5$.
Pascarelle et al. (1996b) have argued that LAEs 
(found at $z\simeq 2.4$) are sub-galactic objects 
from their small sizes.
Those LAEs have also been suggested to be very young 
(Keel et al. 2002).
If LAEs at $z\simeq 5$ are also sub-galactic and 
if LBGs have relatively large stellar masses as suggested by, e.g., 
Papovich, Dickinson, \& Ferguson (2001) and 
Shapley et al. (2001), 
then the high mass of haloes hosting $z\simeq 5$ LAEs 
found here will imply that dark haloes of a given mass 
host galaxies of a wide range of stellar mass 
or a wide range of evolutionary stages.


\acknowledgments
We would like to thank the referee, William Keel, for 
useful comments.
We are grateful to the Subaru Telescope staff
for their invaluable help in our observations.
H. Furusawa, F. Nakata, and M. Ouchi
acknowledge support from the Japan Society for the
Promotion of Science (JSPS) through JSPS Research Fellowships
for Young Scientists.




\clearpage









\begin{thebibliography}{}

\bibitem[Bernardeau (1994)]{bernardeau1994}
Bernardeau, F.,
1994, \apj, 427, 51

\bibitem[Bertin \& Arnouts(1996)]{bertin1996}
Bertin, E.\ \& Arnouts, S.\ 1996, \aaps, 117, 393

\bibitem[Bruzual & Charlot (1993)]{bruzual1993} 
Bruzual A., G. \& Charlot, S.,
1993, \apj, 405, 538

\bibitem[Campos et al. (1999)]{campos1999} 
Campos, A., Yahil, A., Windhorst, R. A., Richards, E. A., 
Pascarelle, S., Impey, C., \& Petry, C.,
1999, \apj, 511, L1

\bibitem[Coleman et al. (1980)]{coleman1980} 
Coleman, G. D., Wu, C.-C., \& Weedman, D. W., 
1980, \apjs, 43, 393

\bibitem[Dey et al.(1998)]{dey1998}
Dey, A., Spinrad, H., Stern, D., Graham, J. R., \& Chaffee, F.,
1998, \apj, 498, L93

\bibitem[Eke et al.(1996)]{eke1996}
Eke, V. R., Cole, S., \& Frenk, C. S.,
1996, \mnras, 282, 263

\bibitem[Francis et al.(1996)]{francis1996}
Francis, P. J., et al.,
1996, \apj, 457, 490

\bibitem[Geller & Huchra (1989)]{geller1989}
Geller, M. J. \& Huchra, J. P.,
1989, {\it Science}, 246, 897

\bibitem[Giavalisco et al. (1994)]{giavalisco1994}
Giavalisco, M., Steidel, C. C., \& Szalay, A. S.,
1994, \apj, 425, L5

\bibitem[GS (1983)]{gs1983}
Gunn, G. E. \& Stryker, L. L.
1983, \apjs, 52, 121

\bibitem[Kashikawa et al. (2002)]{kashikawa2002}
Kashikawa, N. et al., 2002, \pasj, in press

\bibitem[Keel et al. (1999)]{keel1999}
Keel, W. C., Cohen, S. H., Windhorst, R. A., 
\& Waddington, I., 
1999, \aj, 118, 2547

\bibitem[Keel et al. (2002)]{keel2002}
Keel, W. C., Wu, W., Waddington, I., Windhorst, R. A., 
\& Pascarelle, S. M.,
2002, \aj, 123, 3041

\bibitem[Kinney et al.(1996)]{kinney1996} 
Kinney, A. L., Calzetti, D., Bohlin, R. C., McQuade, K., 
Storchi-Bergmann, T., \& Schmitt, H. R.
1996, \apj, 467, 38

\bibitem[LeFevre et al. 1996]{lefevre1996}
Le F\`evre, O., Deltorn, J. M., Crampton, D., \& Dickinson, M.,
1996, \apj, 471, L11

\bibitem[Madau 1995]{madau1995}
Madau, P.
1995, \apj, 441, 18

\bibitem[Maihara et al. 2001]{maihara2001}
Maihara, T. et al.,
2001, \pasj, 53, 25

\bibitem[Malkan et al. 1996]{malkan1996}
Malkan, M. A., Teplitz, H., \& McLean, I. S., 
1996, \apj, 468, L9

\bibitem[Miyazaki et al.(2002)] {miyazaki2002}
Miyazaki, S. et al. 2002, \pasj, in press

\bibitem[Mo & White (1996)]{mo1996}
Mo, H. J. \& White, S. D. M.,
1996, \mnras, 282, 347

\bibitem[Ouchi et al. (2001)]{ouchi2001}
Ouchi, M. et al., 2001, \apj, 558, L83

\bibitem[Ouchi et al. (2003a)]{ouchi2003a}
Ouchi, M. et al., 2003a, \apj, 582, 60

\bibitem[Ouchi et al. (2003b)]{ouchi2003b}
Ouchi, M. et al., 2003b, in preparation

\bibitem[Pascarelle et al. (1996a)]{pascarelle1996a}
Pascarelle, S. M., Windhorst, R. A., Driver, S. P., 
Ostrander, E. J., \& Keel, W. C., 
1996a, \apj, 456, L21

\bibitem[Pascarelle et al. (1996b)]{pascarelle1996b}
Pascarelle, S. M., Windhorst, R. A., Keel, W. C., \& Odewahn, S. C.,
1996b, \nat, 383, 45

\bibitem[Papovich et al. (2001)]{papovich2001}
Papovich, C., Dickinson, M., \& Ferguson, H.,
2001, \apj, 559, 620

\bibitem[Pentericci et al. (2000)]{pentericc2000}
Pentericci, L. et al.,
2000, \aap, 361, L25

\bibitem[Porciani & Giavalisco (2002)]{porciani2002}
Porciani, C. \& Giavalisco, M.,
2002, \apj, 565, 24

\bibitem[Shapley et al. (2001)]{shapley2001}
Shapley, A. E., Steidel, C. C., Adelberger, K. L., 
Dickinson, M., Giavalisco, M., \& Pettini, M.
2001, \apj, 562, 95

\bibitem[Steidel et al. (1998)]{steidel1998}
Steidel, C. C., Adelberger, K. L., Dickinson, M., Giavalisco, M., 
Pettini, M., \& Kellogg, M., 
1998, \apj, 492, 428


\bibitem[Venemans et al. (2002)]{venemans}
Venemans, B. P. et al.,
2002, \apj, 569, L11

\bibitem[Yagi et al. (2002)]{Yagi2002}
Yagi, M., Kashikawa, N., Sekiguchi, M., Doi, M., Yasuda, N., 
Shimasaku, K., \& Okamura, S.
2002, \aj, 123, 66

\end{thebibliography}
\end{document}